\documentstyle[12pt]{article}

\advance\textheight by \topskip
\begin{document}
\tolerance=100000
\thispagestyle{empty}
\setcounter{page}{0}

\newcommand{\be}{\begin{equation}}
\newcommand{\ee}{\end{equation}}
\newcommand{\br}{\begin{eqnarray}}
\newcommand{\er}{\end{eqnarray}}
\newcommand{\ba}{\begin{array}}
\newcommand{\ea}{\end{array}}
\newcommand{\bi}{\begin{itemize}}
\newcommand{\ei}{\end{itemize}}
\newcommand{\bn}{\begin{enumerate}}
\newcommand{\en}{\end{enumerate}}
\newcommand{\bc}{\begin{center}}
\newcommand{\ec}{\end{center}}
\newcommand{\ul}{\underline}
\newcommand{\ol}{\overline}
\newcommand{\ttb}{$t\bar t$}
\newcommand{\bbb}{$b\bar b$}
\newcommand{\epem}{$e^+e^-$}
\newcommand{\eeqq}{$e^+e^-\rightarrow q\bar q$}
\newcommand{\eett}{$e^+e^-\rightarrow t\bar t$}
\newcommand{\eebb}{$e^+e^-\rightarrow b\bar b$}
\newcommand{\ttbbww}{$t\bar t \rightarrow b\bar b W^+W^-$}
\newcommand{\eettbbww}{$e^+e^-\rightarrow t\bar t \rightarrow b\bar b W^+W^-$}
\newcommand{\eebbww}{$e^+e^-\rightarrow b\bar b W^+W^-$}
\newcommand{\bbww}{$b\bar b W^+W^-$}
\newcommand{\bbbar}{$ b\bar b$}
\newcommand{\ttbar}{$ t\bar t$}
\newcommand{\ar}{\rightarrow}
\newcommand{\sm}{${\cal {SM}}$}
\newcommand{\Dir}{\kern -6.4pt\Big{/}}
\newcommand{\Dirin}{\kern -10.4pt\Big{/}\kern 4.4pt}
\newcommand{\DDir}{\kern -7.6pt\Big{/}}
\newcommand{\DGir}{\kern -6.0pt\Big{/}}

\def\Ord{\buildrel{\scriptscriptstyle <}\over{\scriptscriptstyle\sim}}
\def\OOrd{\buildrel{\scriptscriptstyle >}\over{\scriptscriptstyle\sim}}
\def\pl #1 #2 #3 {{\it Phys.~Lett.} {\bf#1} (#2) #3}
\def\np #1 #2 #3 {{\it Nucl.~Phys.} {\bf#1} (#2) #3}
\def\zp #1 #2 #3 {{\it Z.~Phys.} {\bf#1} (#2) #3}
\def\pr #1 #2 #3 {{\it Phys.~Rev.} {\bf#1} (#2) #3}
\def\prep #1 #2 #3 {{\it Phys.~Rep.} {\bf#1} (#2) #3}
\def\prl #1 #2 #3 {{\it Phys.~Rev.~Lett.} {\bf#1} (#2) #3}
\def\mpl #1 #2 #3 {{\it Mod.~Phys.~Lett.} {\bf#1} (#2) #3}
\def\rmp #1 #2 #3 {{\it Rev. Mod. Phys.} {\bf#1} (#2) #3}
\def\sjnp #1 #2 #3 {{\it Sov. J. Nucl. Phys.} {\bf#1} (#2) #3}
\def\cpc #1 #2 #3 {{\it Comp. Phys. Comm.} {\bf#1} (#2) #3}
\def\xx #1 #2 #3 {{\bf#1}, (#2) #3}
\def\preprint{{\it preprint}}

\begin{flushright}
{ DFTT 03/95}\\
{ DTP/95/08}\\
{ Cavendish--HEP--95/14}\\
{ October 1995\hspace*{.5 truecm}}\\
\end{flushright}

\vspace*{\fill}

\begin{center}
{\Large \bf
\eebbww\ events at the Next Linear Collider: colour structure of top
signal and irreducible background}\\[1.5cm]
{\large A.~Ballestrero$^a$, V.A.~Khoze$^b$, E.~Maina$^a$,}\\[0.25 cm]
{\large S.~Moretti$^{c}$ and W.J.~Stirling$^{b,d}$}\\[0.4 cm]
{\it a) Dipartimento di Fisica Teorica, Universit\`a di Torino,}\\
{\it and I.N.F.N., Sezione di Torino,}\\
{\it Via Pietro Giuria 1, 10125 Torino, Italy.}\\[0.25cm]
{\it b) Department of Physics, University of Durham,}\\
{\it South Road, Durham DH1 3LE, United Kingdom.}\\[0.25cm]
{\it c) Cavendish Laboratory, University of Cambridge,}\\
{\it Madingley Road, Cambridge CB3 0HE, UK.}\\[0.25cm]
{\it d) Department of Mathematical Sciences, University of Durham,}\\
{\it South Road, Durham DH1 3LE, United Kingdom.}\\[0.5cm]
\end{center}

\vspace*{\fill}

\begin{abstract}
\noindent We examine the colour
structure and charged particle yield for both the $t \bar t$ signal and
the irreducible background processes contributing to
\eebbww\ production close to the \ttb\ threshold.
The charged particle multiplicity for the various components of
the cross section is computed as a function of several kinematic variables.
Our study may have important implications for recently proposed
studies of interconnection phenomena in \ttb\  production at high--energy
$e^+e^-$ colliders.
\end{abstract}

\vspace*{\fill}
\newpage

\subsection*{1. Introduction}

One of the most important physics topics at future \epem\ linear
colliders is the detailed study of the top quark. In particular,
a unique precision in the top mass determination (with an accuracy
of a few hundred MeV) is anticipated, see for example
Refs.~\cite{NLC,ee500,ppp,LC92}.
One of the obvious requirements for the  success of such precise studies is
a detailed understanding of
the background processes, in particular of the colour structure of the
\eebbww\ background events.

Recall that the dominance of the $t\ar bW^+$ weak decay mode leads
to a large top  width
$\Gamma_t$, which is approximately 1.6 GeV for
 $m_t\simeq 180$ GeV.
This width is larger than the typical hadronic scale $\mu\sim 1$ fm$^{-1}$,
and  so the top decays before it has time to hadronize \cite{kuhn,bigi}.
It is precisely this large decay width that makes top physics so unique.
First, the top decay width $\Gamma_t$ provides an infrared
cut--off for the strong forces between the $t$ and the $\bar t$
\cite{fadin1,fadin2,fadin3}: the width acts as a physical
`smearing'
\cite{poggio}, and top production becomes a quantitative prediction of
perturbative QCD, largely independent of non--perturbative phenomenological
algorithms. Second, $\Gamma_t$ controls the size of QCD interferences between
radiation occurring at different stages of the \ttb\ production process
\cite{khoze1,yu,khoze2}. These interferences affect the structure of the
colour flows in the \ttb\
events, and may provide a potentially serious source of uncertainty
in the reconstruction of the final state.

As is discussed in detail in Ref.~\cite{ks} (see also \cite{vak}), particle
production in events of the type
\be\label{eettbbww}
e^+e^-\rightarrow t\bar t \rightarrow b\bar b W^+W^-
\ee
could depend in a non--trivial way on the kinematics of the process.
The effects of {\it energetic} perturbative gluon radiation
are suppressed because of the space--time separation between the decays
of the $t$ and $\bar t$ quarks.
Soft gluon emission and non--perturbative fragmentation introduce a
correlation (interconnection) between the $b$-- and $\bar b$--jets coming
from the top decays. As a result, the $b$ and $\bar b$ have to
`cross--talk' in order to produce a final state made up of
colourless hadrons\footnote{Analogously to Ref.~\cite{ks}, throughout this
paper
we assume for simplicity that the $W$'s decay leptonically.}.

Such interconnection phenomena could affect the final state in \ttb\ events
 in many respects.
In order to  quantify their size, it was proposed in Ref.~\cite{ks}
to examine the average multiplicity of double leptonic
\ttb\ decays as a function of the relative angle between the $b$--
and the $\bar b$--jets.
Such studies could provide  important information
about the size of the interconnection--related systematic uncertainties
in the top mass reconstruction. They are also interesting in their own right,
since they potentially open up a new laboratory for probing
non--perturbative QCD
dynamics, see for example Refs.~\cite{ks,vak,gpz,sz}.

The issue of background processes was not addressed in
Ref.~\cite{ks}.  Since the
interconnection effects in the distributions of the final--state particles
in the process
\be\label{eebbww}
e^+e^-\rightarrow b\bar b W^+W^-
\ee
are comparatively weak, and since the background contributions are
quite different in various kinematical regions, the detailed properties
of the colour flow structure of
the background events need special detailed consideration.
Here we present a complete analysis of the full process
(\ref{eebbww})
by studying (i) the detailed colour structure of
\ttb\ production and decay, and (ii) the  colour structure of
the irreducible backgrounds in \bbww\ production.

This paper is organised as follows. In Section 2 we give a detailed
description of the structure of the matrix element for the reaction
\eebbww. Numerical  results are given in Section 3 and  the conclusions
are presented in Section 4.

\subsection*{2. Colour Structure of the Matrix Element}

The matrix element  used in this study is the one already
introduced in Refs.~\cite{eett,eezh}, where further details
(method of calculation, parameter values, etc.) can be found.
Only the value of the  Higgs mass  $M_{H}$ has been changed in the
present work.
Since in the present context
we are not interested in  effects due to virtual
Higgs states, we deliberately increased $M_H$ to a value greater
than the collision energy, viz.
 $M_H = 600$~GeV for $\sqrt s\le 500$~GeV.
% \footnote{In fact, even though we know
%that for $M_H\Ord300$ GeV at $\sqrt s\le 500$ GeV
%the Higgs diagram contributions to the total \eebbww\ cross section is
%not completely negligible \cite{eett,eezh},
%especially when compared to the other resonant
%backgrounds,   nothing prevents us
%from assuming a Higgs with a much larger mass, since no strict upper
%bound exists.}.
We use the value $m_t=175$~GeV for the top quark mass, as suggested
by recent data \cite{discovery}.

%We next  describe all the different colour structures of the
%various components of the signal and background  contributions
%of \eebbww\ production. There are basically four
%different mechanisms \cite{eett,eezh}, illustrated in Fig.~1:
%\begin{itemize}
%\item[{{(i)}}]   double $t$--resonance (i.e. $t\bar t\ar b\bar b W^+W^-$),
%\item[{{(ii)}}]  single $t$--resonance (i.e. $t\ar b W^+$ + charged
%conjugate),
%\item[{{(iii)}}] $Z^0$--resonance (i.e. $Z^0\ar b\bar b$),
%\item[{{(iv)}}] non--resonant (NR), including the splitting
%process $\gamma^*\ar b\bar b$,
%\end{itemize}
%that are obtained by integrating over  phase space
%the terms ${\cal M}_1^2$, $2{\cal M}_2^2$,
%${\cal M}_4^2$ and ${\cal M}_8^2$ (defined in Ref.~\cite{eezh})
%respectively\footnote{The
%factor  two multiplying ${\cal M}_2^2$
%comes from the charge -conjugation invariance relating
%the two single top contributions $t\ar b W^+$ and $\bar t\ar \bar b W^-$.}.

With the assumption that the Higgs mass is
always larger than the collider CM energy,
the dominant contributions to the modulus squared of the scattering amplitude
  are  (see Ref.~\cite{eezh}):
\begin{itemize}
\item[{{(i)}}]   double $t$--resonance (i.e. $t\bar t\ar b\bar b W^+W^-$,
Fig.~1a),
\item[{{(ii)}}]  single $t$--resonance (i.e. $t\ar b W^+$ + charged
conjugate,  Fig.~1b),
\item[{{(iii)}}] $Z^0$--resonance (i.e. $Z^0\ar b\bar b$,
Fig.~1c),
\item[{{(iv)}}] non--resonant (NR)  (Fig.~1d together with
those in Fig.~1c involving the splitting
process $\gamma^*\ar b\bar b$).
\end{itemize}

If we denote by  $M_j$ the sum of the diagrams
in the $j$th channel, and  neglect Higgs contributions, we have
\be\label{sum} M_{tot}=\sum_{j=1}^{4} M_j,\ee
where $M_{tot}$ is the total Feynman amplitude.
In squaring Eq.~(\ref{sum}), we define the combinations\footnote{Note
that only the {\it sum} of these is positive definite. In particular,
the interference terms in ${\cal M}_{4}$ can cause the corresponding
contribution to the cross section to be zero or negative, see Figs.~3 and 4
below.}
\begin{eqnarray}
\label{m1} {\cal M}_{1}&=&|M_1|^2,                       \\
\label{m2} {\cal M}_{2}&=&|M_2|^2+2\mbox{Re}[M_1M_2^*]    \\
\label{m7} {\cal M}_{3}&=&|M_3|^2,                       \\
\label{m8} {\cal M}_{4}&=&|M_4|^2
               + \mbox{all remaining interference terms},
\end{eqnarray}
such that
\be |M_{tot}|^2=\sum_{j=1}^{4} {\cal M}_j.\ee
We associate these four combinations with  the contributions (i) -- (iv)
defined above.

 When discussing the structure of the
particle flows corresponding to the background processes (ii), (iii) and (iv),
 it is convenient
to apply the standard `parton shower plus fragmentation' picture for
$e^+e^-\ar \gamma^*,Z^0\ar q\bar q$, various aspects of which are now
well understood. Analogously to Ref.~\cite{ks}, the colour flow
distribution can easily
be described using the language of  QCD antennae/dipoles
(for details see \cite{book}).
Thus, the radiation pattern corresponding to the diagrams
of Fig.~1c and 1d is given by the $\widehat{b\bar b}$ antenna which describes
the production of hadrons in the process $\gamma^*,Z^0\ar b\bar b$.
Because of the large amount of data from PETRA/PEP and LEP1/SLC,
the properties of $b$--jets in this process are now well understood.
A specific role is played here by kinematical configurations
in which  the $b$-- and $\bar b$--jets come from the decay of an
 on--mass--shell
$Z^0$ boson. For these configurations the average particle multiplicity
 is given by the
multiplicity in $Z^0\ar b\bar b$, as measured for example
at LEP1.
For the non--resonant processes in Fig.~1 the average
multiplicity of the final--state particles  depends strongly
on the relative
orientation of the $b$-- and $\bar b$--jets. In contrast, for the double
$t$--resonance (Fig.~1a) and single $t$--resonance (Fig.~1b)
mechanisms the angular dependence
is comparatively weak (but nonetheless readily visible)
 because of the suppressed
role of the $\widehat{b\bar b}$ antennae \cite{ks}.

It is worthwhile to mention that not far from the \ttb\ threshold the
final--state structure corresponding to  double top--resonant production
is dominated by two essentially azimuthally
symmetric $b$-- and $\bar b$--jets,
while the background $\gamma^*,Z^0\ar b\bar b$
processes are described by the $\widehat{b\bar b}$ antenna
patterns with all the appropriate azimuthal asymmetry effects
(see Ref.~\cite{book}). Moreover,
the $\gamma^*\ar b\bar b$ contribution enhances the role
of low multiplicity events. Note also that with increasing collision energy the
$b$-- and $\bar b$--jets coming from  \ttb\ production tend to be more
back--to--back, while those resulting from $\gamma^*,Z^0\ar b\bar b$
splitting tend  to become more collinear.

\subsection*{3. Results}

Our numerical results are presented in  Figs.~2--8.
In all of them we distinguish between the   \eebbww\
components  (i) --- (iv) discussed in the previous section
 and in Refs.~\cite{eett,eezh}.

In Fig.~2 we show the dependence of the various contributions
on the collider centre-of-mass (CM) energy,
in the range $200~{\rm GeV}\Ord\sqrt s\Ord 500~{\rm GeV}$.
The behaviour of the various curves can be readily understood in terms
of the diagrams contributing to the ${\cal M}_i^2$ amplitudes, see Fig.~1.
The $t\bar t\ar b\bar b W^+W^-$ curve corresponds to
\ttb--pair production followed by the decays
$t\ar b W^+$ and $\bar t\ar \bar b W^-$, both above ($\sqrt s\ge350$ GeV) and
below threshold ($\sqrt s<350$ GeV). In contrast to the naive case of the
`narrow width approximation' (see Ref.~\cite{eett}), in which
this process is computed in terms of a {\it production$\times$ decay} reaction
$\sigma(e^+e^-\ar t\bar t)\times [BR(t\ar bW)]^2$ with zero
top width in the production process (top quarks  produced
on--shell) and independent semileptonic top decays, here both $t$--quark finite
width effects  and energy/spin correlations between
the two  top decays are included. The consequence\footnote{Other effects
are discussed in Ref.~\cite{eett}. Note that analogously to Ref.~\cite{eett}
we do not address here the issue of the QCD Coulomb
attraction effects in process (\ref{eettbbww}).
These effects are especially
important near the \ttb\ threshold where they substantially enhance the
production
amplitude. The Coulomb physics could be incorporated using the
technique of  non--relativistic Green's functions for the
\ttb\ system \cite{fadin1,fadin2} which  has now become quite routine, see
Refs.~\cite{NLC,ee500,ppp,LC92}
and references therein. The background amplitudes
which are our main concern in this paper are only weakly affected
by  QCD final--state interactions. We should also mention
that in the present study we do not take into account  QED Initial
State Radiation (ISR), but these effects would be straightforward to include.}
of this is clearly
visible in the dependence of the cross section on $\sqrt{s}$.
In particular, there is a significant
 `shoulder' which extends far below the nominal threshold.

This behaviour naturally also affects the other \eebbww\ contributions that
include interferences with the double top resonance channel, such as
$t\ar b W^+$ and NR. This is clearly visible in both of these as
 the little `bump'  around $\sqrt s=2m_t$, which slightly decreases
the cross section in the first case and enhances it in the second.
It is important to stress that these curves,
and therefore their interplay, have no {\it separate} physical
meaning,  since  the ${\cal M}_i^2$'s  are not separately gauge invariant,
but they provide  a useful way of
looking inside the process and distinguishing between the different
interactions. The final step must always  be to sum
together all the various components ${\cal M}_i^2$ and consider the total
rates.

On their own, the curves in Fig.~2 corresponding to $t\ar b W^+$ and NR
describe, respectively,  the production of a single top quark
through a highly virtual
$W^{\pm*}$ decaying to $tb$ pairs with $t\ar bW$,
and dominantly (as  will be apparent from Figs.~3--4, see below)
of a $b\bar b$--pair from $\gamma^*\ar b\bar b$. In the first case
 the main contribution
to  single top production comes from diagrams in which the decaying
$W^\pm$ is produced via $t$--channel exchange (i.e. the second diagram
of Fig.~1b) rather than
via $s$--channel exchange (i.e. the first and third diagrams of Fig.~1b),
since the corresponding
curve in Fig.~2 does not show any decrease with $\sqrt s$ above the
nominal  threshold
at $ m_t+M_{W}+m_b \approx 260$~GeV.
In the second case, except for the range $270~{\rm GeV}\Ord \sqrt s\Ord 360$
GeV where  the  interferences between  non--resonant and resonant
diagrams are important, the shape is almost flat, reflecting the
dominance of the  $t$--channel contribution
(the first, second and fifth diagrams of Fig.~1c, with a $\gamma$  propagator)
to $\gamma^*W^+W^-$ production followed
by the $\gamma^*\ar b\bar b$ splitting.
Finally, the curve corresponding to the $Z^0\ar b\bar b$ background illustrates
the characteristic features of the production of three vector bosons
 $Z^{0} W^+W^-$,
with the $Z^0$ decaying to the $b\bar b$--pair, with a threshold at
$2M_{W}+M_{Z}\approx 250$~GeV and a dominant $t$--channel exchange
structure (the first, second and fifth diagrams of Fig.~1c, with a
$Z^0$--propagator).

As already  mentioned, we concentrate on the case
of leptonic decays of {\it both} the $W^\pm$'s. This implies that we have
final states with two neutrinos  and therefore with
 missing energy and transverse
momentum:
$b\bar b W^+W^-\ar b\bar b (\ell\nu_\ell)(\ell'\nu_{\ell'})$ or
more generally
$jj W^+W^-\ar jj (\ell\nu_\ell)(\ell'\nu_{\ell'})$, depending on whether
 or not one can  exploit the possibility of using $b$--tagging
\cite{tag}.
This missing energy and  transverse
momentum prevents us from studying
the kinematics of \eebbww\ events by looking at variables which require
 the reconstruction of the $W^\pm$'s through their decay products,
 for example  the invariant masses
$M_{bW}$ or  energies $E_{bW}$. These quantities
would of course be
extremely useful in distinguishing  the $t\bar t$  signal from
 the other \eebbww\ background contributions
 \cite{eett,eezh}.
Thus in the figures which follow we restrict our attention to
variables involving the $b\bar b$--system only.

In Fig.~3 we show the dependence of the four components { (i)--(iv)}
on the cosine of the angle between  the $b$ and $\bar b$, at $\sqrt s=360$~GeV.
The convenience of this variable has been  explained
above (see also  Ref.~\cite{ks}).
%\footnote{Note that $\theta_{b\bar b}$ is not, in principle, a quantity
%which is accessible directly in the experiment (for a discussion of the
% application to a realistic  experimental analysis see
%Ref.~\cite{ks}).}
 For  double and single top production the angular dependence is weak,
 reflecting the absence of any strong correlation between the top
  and anti-top decay products near threshold.
In contrast,  the  $Z^0\ar b\bar b$ and NR components exhibit
a more pronounced structure.
For the former, we note that there  is a significant boost
to the $Z^0$  at this centre--of--mass energy (note that $\sqrt s=360$~GeV
is 110~GeV above the $Z^0W^+W^-$ threshold), and so the peak at
$\cos(b\bar b) = -1$ corresponding to back--to--back production in the
$Z^0$ rest frame is significantly smeared out.
For the NR component, the main feature is the $\gamma^*\ar b\bar b$
contribution which prefers small invariant masses and therefore has
$\cos(b\bar b) \approx 1$.

In Figs.~4 and 5 we show the dependence of the four components
 on the invariant mass of
the $b\bar b$--system and on the energy of the $b(\bar b)$ respectively.
As expected, for the former we can clearly recognize  both the
$M_{b\bar b} \approx M_{Z}$ peak in the $Z^0\ar b\bar b$ component  and  the
$\gamma^*\ar b\bar b$ contribution at small $M_{b\bar b}$ in NR.
Neither the double top nor the single top channel show any particularly
distinctive features.  Fig.~5 indicates
the average energy of the $b$'s coming from the various subprocesses:
approximately $70$~GeV for $t\bar t\ar b\bar b W^+W^-$,  80~GeV
for $t\bar bW^+$  and $60$~GeV for $Z^0\ar b\bar b$.
For the NR component there is a preference for small $b$ energies.

%%%%%%%%%%%%%%%%%%%%%

We next adopt a semi--qualitative procedure for evaluating
the average charged multiplicities corresponding to the different contributions
to the \eebbww\ events not far from the \ttb\ threshold. This procedure is
based on the antenna pattern analysis \cite{khoze1,yu,khoze2} of soft gluon
radiation in  top quark production events within the framework of the
MLLA--LPHD picture of  multiple hadron production \cite{book}.

Our starting point is the multiplicity  $N_{b\bar b}(W)$ in \eebb\
production at CM energy
$W=\sqrt s=2E_b$, where $E_b$ is the energy of the $b$--jet in
the CM system. According to the MLLA-LPHD scenario (see
Ref.~\cite{bbbar} for details)
$N_{b\bar b}$ can be expressed in terms of the multiplicity
$N_{q\bar q}$ in the light--quark production process \eeqq\ as
\be\label{mult1}
N_{b\bar b}(W)=N_{q\bar q}(W)+\delta_b,
\ee
where the difference $\delta_b$ is $W$--independent.
This prediction of perturbative QCD is in a good agreement with
existing experimental data \cite{b_data}, from which  one obtains
 $\delta_b \simeq 3$.
For $N_{q\bar q}(W)$ one can use the MLLA--inspired
simple analytical formula proposed in Ref.~\cite{dk}. As a result, the
charged particle multiplicity in \bbb\ events can be readily evaluated
as\footnote{Of course one could alternatively  apply all the powerful
machinery of the  successful Monte Carlo algorithms
(JETSET, HERWIG, ARIADNE).
However for the purposes of this paper we can safely
restrict ourselves to the estimates based on Eq.~(\ref{mult2}).
It is remarkable
that in a wide energy domain the predictions of
Eq.~(\ref{mult2}) are in a very
good agreement with the results of the
JETSET parton shower Monte Carlo program \cite{s}.
(We are  grateful to T.~Sj\"ostrand
for providing us with the numerical results of JETSET for $N_{b\bar b}(W)$).}
\be\label{mult2}
N_{b\bar b}^{Z^0, {\rm{NR}}}(E_{b(\bar b)},\cos(b\bar b))\equiv
N_{b\bar b}^{Z^0, {\rm{NR}}}(M_{b\bar b})
\approx 5.85+0.125\exp[2.317\sqrt{\log(M_{b\bar b}/{\rm{GeV}})}],
\ee
with\footnote{The superscripts $t \bar t$, $t$, $Z^0$ and NR in
Eqs.~(\ref{mult2},\ref{mult3})
refer to the production
mechanisms (i)--(iv) of the $b\bar b$--pair in \eebbww\ events.}
$M_{b\bar b}^2= 2(E_bE_{\bar b}-
|p_b||p_{\bar b}|\cos(b\bar b)+  m_b^2) $  and
$|p_{b(\bar b)}|^2=E_{b(\bar b)}^2-m_b^2$.

Motivated by the results of Refs.~\cite{khoze1,yu,khoze2} we can approximate
the average multiplicities corresponding to the  $t$--resonant production
processes ((i) and (ii)) by
\be\label{mult3}
N_{b\bar b}^{t\bar t, t}(E_{b(\bar b)},\cos(b\bar b))\equiv
N_{b\bar b}^{t\bar t, t}(E_b,E_{\bar b})
\approx
\frac{1}{2}N^{Z^0, {\rm{NR}}}_{b\bar b}(M_{b\bar b}\ar 2E_b)+
\frac{1}{2}N^{Z^0, {\rm{NR}}}_{b\bar b}(M_{b\bar b}\ar 2E_{\bar b}),
\ee
where $E_b$ and $E_{\bar b}$ are the actual CM energies of the $b$--
and $\bar b$--jets, respectively. This result
 accounts for the contributions
of the antennae of the type $\widehat{tb}$, $\widehat{\bar t\bar b}$ or
$\widehat{t\bar b}$, $\widehat{\bar t b}$
(note that  interconnection effects are not included here).

Using the above formulae (i.e. Eq.~(\ref{mult2}) for
the multiplicity for the $Z^0\ar b\bar b$ and the non--resonant contributions
((iii) and  (iv))   and
Eq.~(\ref{mult3}) for the
double and single $t$--resonant contributions ((i) and (ii)),
we can study  the quantities
\be\label{fig6}
\langle N^i_{\rm{cos}} \rangle =\frac{\int
dE_{b(\bar b)}\;  {\cal{I}}^i(E_{b(\bar b)},\cos(b\bar b))}
                  {\int dE_{b(\bar b)}\; {\cal{J}}^i
                  (E_{b(\bar b)},\cos(b\bar b))},
\ee
\be\label{fig7}
\langle N^i_E
 \rangle=\frac{\int d\cos(b\bar b)\;  {\cal{I}}^i(E_{b(\bar b)},\cos(b\bar b))}
           {\int d\cos(b\bar b)\; {\cal{J}}^i(E_{b(\bar b)},\cos(b\bar b))},
\ee
at fixed CM energy ($\sqrt s=360$ GeV), and
\be\label{fig8}
\langle N^i_{\rm{tot}} \rangle =
\frac
{\int dE_{b(\bar b)} d\cos(b\bar b)\;  {\cal{I}}^i(E_{b(\bar b)},\cos(b\bar
b))}
{\int dE_{b(\bar b)} d\cos(b\bar b)\;
{\cal{J}}^i(E_{b(\bar b)},\cos(b\bar b))},
\ee
as a function of  the CM energy in the range 240--390~GeV, with $i=t\bar
t,t,Z^0$ and NR.
In Eqs.~(\ref{fig6})--(\ref{fig8})
${\cal I}^i$ and ${\cal J}^i$ are given by
\be\label{integrand1}
{\cal{I}}^i(E_{b(\bar b)},\cos(b\bar b))=\frac{d\sigma^i}{dE_{b(\bar b)}
d\cos(b\bar b)}\ N^i_{b\bar b}(E_{b(\bar b)},\cos(b\bar b)),
\ee
and
\be\label{integrand2}
{\cal{J}}^i(E_{b(\bar b)},\cos(b\bar b))=\frac{d\sigma^i}{dE_{b(\bar b)}
d\cos(b\bar b)},
\ee
where $\sigma^i$ is the total `cross section'
of the $i$--th component.

The above quantities can be interpreted as the average number of
charged particles   per unit interval of
the cosine of the angle between the $b$--
and $\bar b$--jets (Eq.~(\ref{fig6})) and  per unit interval
of the energy of the $b(\bar b)$--jets (Eq.~(\ref{fig7})),
and as the total number of charged particles for all  $\cos(b\bar b)$
angles and $E(b)$ energies (Eq.~(\ref{fig8})).
These are defined for each
of the four components {(i)--(iv)} separately.
The distributions for $\langle N^i_{\rm{cos}} \rangle$,
$ \langle N^i_E \rangle $ and $ \langle N^i_{\rm{tot}} \rangle$
are plotted in Figs.~6, 7 and 8,
respectively\footnote{Note that
in Figs.~6--8 a cut on the energy of the $b$-- and $\bar b$--jets has
been applied, i.e. $E(b,\bar b)>20$ GeV.}.

In Fig.~6 the multiplicity arising from the double and
single $t$--resonant contributions  is essentially independent of the angle
between the $b$--jets. This follows from the fact that (i) the multiplicity
here (Eq.~(\ref{mult3})) does not depend explicitly on the angle, and (ii)
the distribution of the energy of the $b$--quarks (Fig.~4) is rather
sharply peaked in both cases around $60 - 80$~GeV. The multiplicity
from the $Z^0\ar b\bar b$  contribution is also approximately
independent of the angle (reflecting the fact that $M_{b\bar b} \approx M_Z$
is constant) apart from near
$\cos(b\bar b) = 1$ where $M_{b\bar b} < M_Z$. In this region, however, the
corresponding cross section is  very small, see Fig.~3. The non--resonant
contribution decreases steadily as $\cos(b\bar b)$
increases as the $\gamma^* \to b \bar b$ contribution (which
prefers small $M_{b\bar b}$ and hence smaller multiplicity)
 becomes more important.

In Fig.~7 the multiplicity arising from the double and
single $t$--resonant contributions rises smoothly with increasing $E_b$,
as expected from Eq.~(\ref{mult3}). The slight fall off at very large
$E_b$ (where the cross section is very small, see Fig.~5) is caused
by asymmetric configurations where $E_b \gg E_{\bar b}$ or {\it vice versa}.
The multiplicity
from the $Z^0\ar b\bar b$  contribution is essentially constant over the whole
energy range and the non--resonant contribution multiplicity
increases quite sharply
with $E_b$, reflecting the underlying dependence on $M_{b\bar b}$.

The collider energy dependence of the multiplicities arising from the
various contributions is shown in Fig.~8. The double and
single $t$--resonant contributions increase
approximately linearly with $\sqrt{s}$, reflecting the underlying
increase in the energy of the $b$--quarks.
The $Z^0\ar b\bar b$  contribution is approximately independent of
$\sqrt{s}$ above the threshold at $2M_W + M_Z \approx 250$~GeV
for producing on-shell $Z^0$ bosons. The non-resonant multiplicity
is somewhat smaller, reflecting the on-average smaller values of
$M_{b \bar b}$, and increases slowly with $\sqrt{s}$.
Also shown (continuous line) in Fig.~8 is  the sum over the four
components,  obtained using  the formula
\be\label{total}
\langle N^{\rm{sum}}_{\rm{tot}}\rangle =\frac
{\sum_i \langle  N_{\rm{tot}}^i\rangle \sigma^i}{\sum_i \sigma^i},
\ee
where $i$ refers as usual to the contributions $t\bar t\ar b\bar b
W^+W^-$, $t\ar bW^+$ + charged conjugate, $Z^0\ar b\bar b$ and non--resonant.
This total multiplicity simply interpolates between the non-resonant
contribution, which dominates the total cross section for
$\sqrt{s} \Ord 250$~GeV, and the double-resonant $t \bar t$
contribution, which dominates the total cross section from just below
the $t \bar t$ threshold, see Fig.~2.

\subsection*{4. Conclusions}

A high--energy $e^+e^-$ collider would provide the ideal
laboratory for precision top--quark physics.
Whether one considers an accurate measurement of the top--quark mass or
a detailed study of
the non--perturbative QCD dynamics and
interconnection effects associated with $t \bar t$ production,
one certainly needs a more detailed understanding
of the colour structure of the background events in order to
separate their influence from the subtle interference phenomena
in the top signal.

In this paper we have examined the colour
structure and charged particle yield in both the $t \bar t$ signal and
background processes contributing to
\eebbww\ production not far from the \ttb\ threshold.
Since the relative contribution of
the background processes to the total particle yield in
\eebbww\ events is relatively small (see Fig. 2),
we have been able to use an approximate procedure for
evaluating the charged particle multiplicities.

Figures~6--8 contain the main results of our study. They show
the multiplicities associated with the various resonant and non--resonant
 components of the cross section, as functions of the energies of the
$b$-- and $\bar b$--jets, the angle between them, and the overall collider
energy. The colour structure of the various components is very different
(see Fig.~1) and this is reflected in the multiplicity distributions.
It is important, therefore, that the background contributions are carefully
taken into account in studies of interconnection phenomena in \ttb\
production.

\subsection*{Acknowledgements}

We thank
T.~Sj\"ostrand for valuable discussions. This work is supported in part by the
Ministero dell' Universit\`a e della Ricerca Scientifica, the UK PPARC,
and   the EC Programme
``Human Capital and Mobility'', Network ``Physics at High Energy
Colliders'', contract CHRX-CT93-0357 (DG 12 COMA).

\goodbreak

\vfill
\newpage

\section*{Figure Captions}

\begin{itemize}

\item[{[1]}] Typical Feynman diagrams describing
          at tree--level the four \eebbww\ components:
          {(i)}   double $t$--resonance (a),
          {(ii)}  single $t$--resonance (b),
          {(iii)} $Z^0$--resonance (c, with $Z^0$--propagators), and
          {(iv)} no resonance (c, with $\gamma$--propagators, and d).
          Internal wavy lines represent  $\gamma$,  $Z^0$ or  $W^\pm$,
          as appropriate.
          Permutations of $(W^+,W^-,\gamma^*/Z^{0*})$ and
          $(\gamma^*,Z^{0*})$ along the fermion lines,
          exchanges between $W^+\leftrightarrow W^-$
          in graphs involving three--vector--vertices, as well as
          the charged conjugate diagrams of (b) are not shown.
%          For their combinations entering in the modulus
%          squared, see Ref.~\cite{eezh} and the text.

\item[{[2]}] Dependence of the individual ``total" cross sections
$\sigma_i$
                 on the collider energy $\sqrt s$
                 for the four \eebbww\ components:
                 {(i)}   double $t$--resonance (continuous line),
                 {(ii)}  single $t$--resonance (dashed line),
                 {(iii)} $Z^0$--resonance (dotted line), and
                 {(iv)} no resonance (chain--dashed line),
                 for $m_t=175$ GeV and $M_{H}=600$ GeV.

\item[{[3]}] Differential distributions $d\sigma_i/d\cos(b\bar b)$
                 in the cosine of the angle between the $b$ and $\bar b$
quarks
                 for the four \eebbww\ components:
                 {(i)}   double $t$--resonance (continuous line),
                 {(ii)}  single $t$--resonance (dashed line),
                 {(iii)} $Z^0$--resonance (dotted line), and
                 {(iv)} no resonance (chain--dashed line), at $\sqrt
                 s=360$ GeV, for $m_t=175$ GeV and $M_{H}=600$ GeV.

\item[{[4]}] Differential distributions $d\sigma_i/dM(b\bar b)$
                 in the invariant mass of the $b\bar b$--pair
                 for the four \eebbww\ components:
                 {(i)}   double $t$--resonance (continuous line),
                 {(ii)}  single $t$--resonance (dashed line),
                 {(iii)} $Z^0$--resonance (dotted line), and
                 {(iv)} no resonance (chain--dashed line), at $\sqrt
                 s=360$ GeV, for $m_t=175$ GeV and $M_{H}=600$ GeV.

\item[{[5]}] Differential distributions $d\sigma_i/dE(b)$
                 in the energy of the $b(\bar b)$ quark
                 for the four \eebbww\ components:
                 {(i)}   double $t$--resonance (continuous line),
                 {(ii)}  single $t$--resonance (dashed line),
                 {(iii)} $Z^0$--resonance (dotted line), and
                 {(iv)} no resonance (chain--dashed line), at $\sqrt
                 s=360$ GeV, for $m_t=175$ GeV and $M_{H}=600$ GeV.

\item[{[6]}] Dependence of $\langle N_{cos}\rangle $, defined in the text,
                 on the cosine of the angle between the $b$ and $\bar b$
                 quarks  for the four \eebbww\ components:
                 {(i)}   double $t$--resonance (continuous line),
                 {(ii)}  single $t$--resonance (dashed line),
                 {(iii)} $Z^0$--resonance (dotted line), and
                 {(iv)} no resonance (chain--dashed line), at $\sqrt
                 s=360$ GeV, for $m_t=175$ GeV and $M_{H}=600$ GeV, after
                 imposing the
                 cut $E(b,\bar b)>20$ GeV on the energy of the $b$'s.

\item[{[7]}] Dependence of $\langle N_E\rangle $, defined in the text,
                 on the energy of the $b(\bar b)$
                 for the four \eebbww\ components:
                 {(i)}   double $t$--resonance (continuous line),
                 {(ii)}  single $t$--resonance (dashed line),
                 {(iii)} $Z^0$--resonance (dotted line), and
                 {(iv)} no resonance (chain--dashed line), at $\sqrt
                 s=360$ GeV, for $m_t=175$ GeV and $M_{H}=600$ GeV, after the
                 cut $E(b,\bar b)>20$ GeV on the energy of the $b$'s.

\item[{[8]}] Dependence of $\langle N_{\rm{tot}}\rangle $, defined in
             the text,
                 on the collider energy $\sqrt s$
                 for the four \eebbww\ components:
                 {(i)}   double $t$--resonance (continuous line),
                 {(ii)}  single $t$--resonance (dashed line),
                 {(iii)} $Z^0$--resonance (dotted line), and
                 {(iv)} no resonance (chain--dashed line), and for
                 their sum (see Eq.~(\ref{total})),
                 for $m_t=175$ GeV and $M_{H}=600$ GeV, after imposing the
                 cut $E(b,\bar b)>20$ GeV on the energy of the $b$'s.

\end{itemize}

\vfill

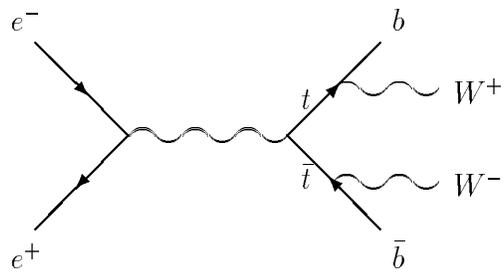
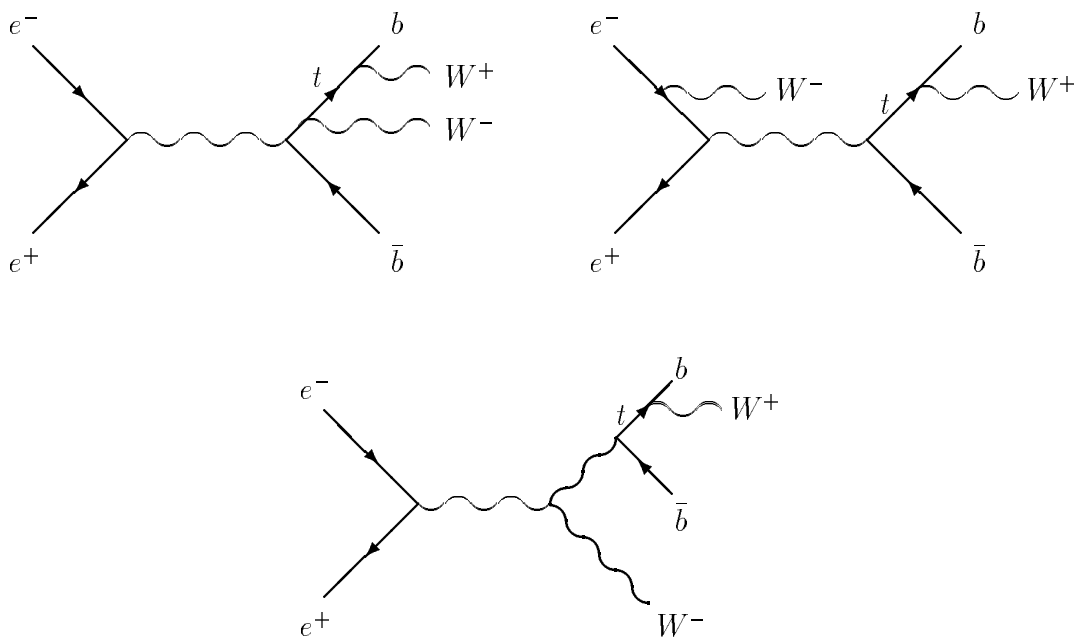
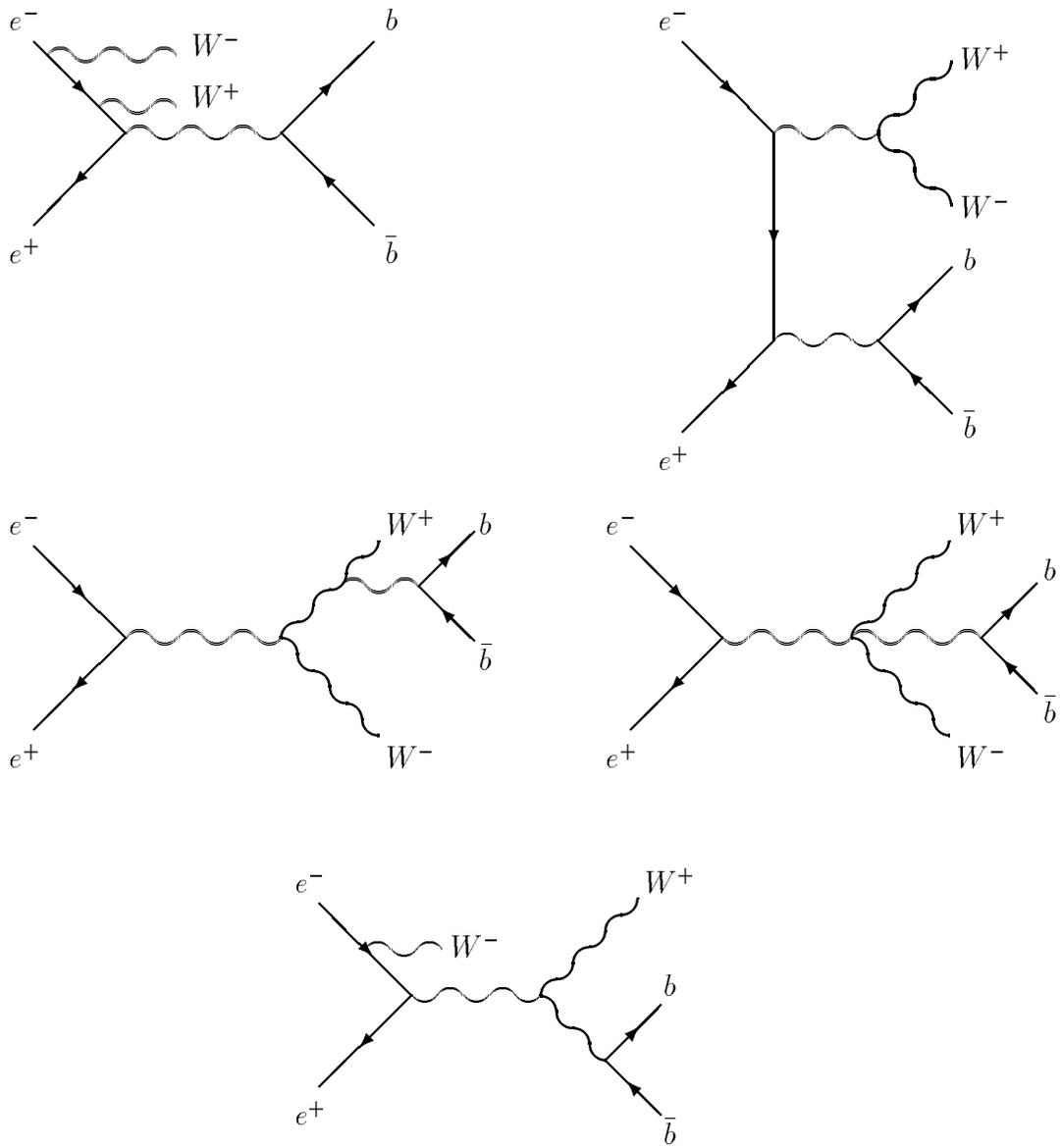
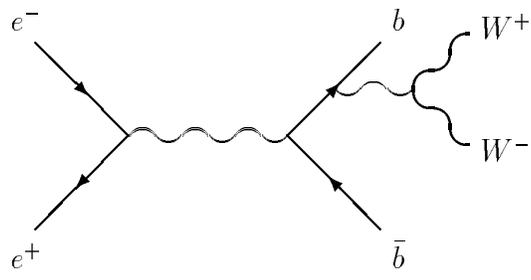
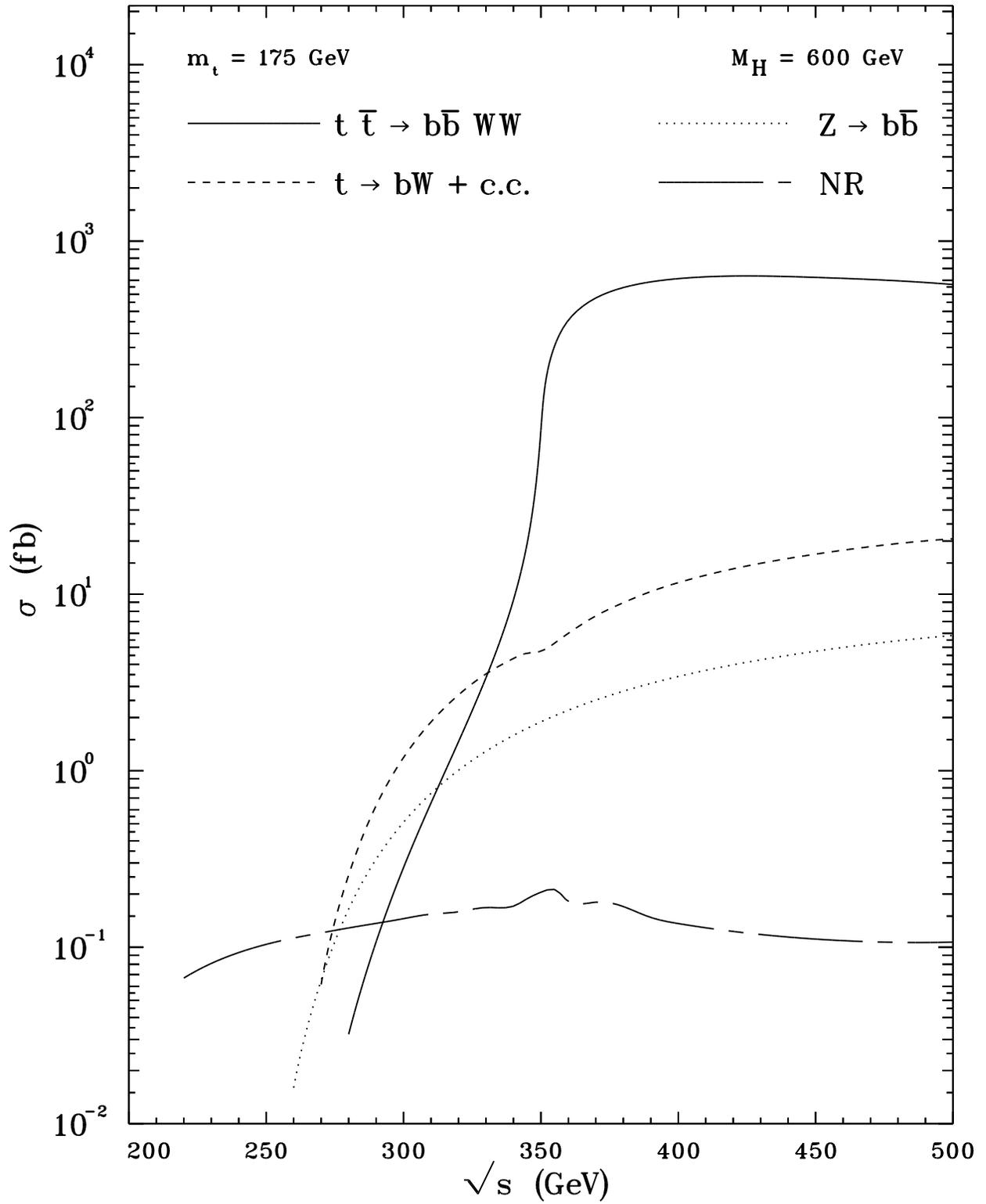
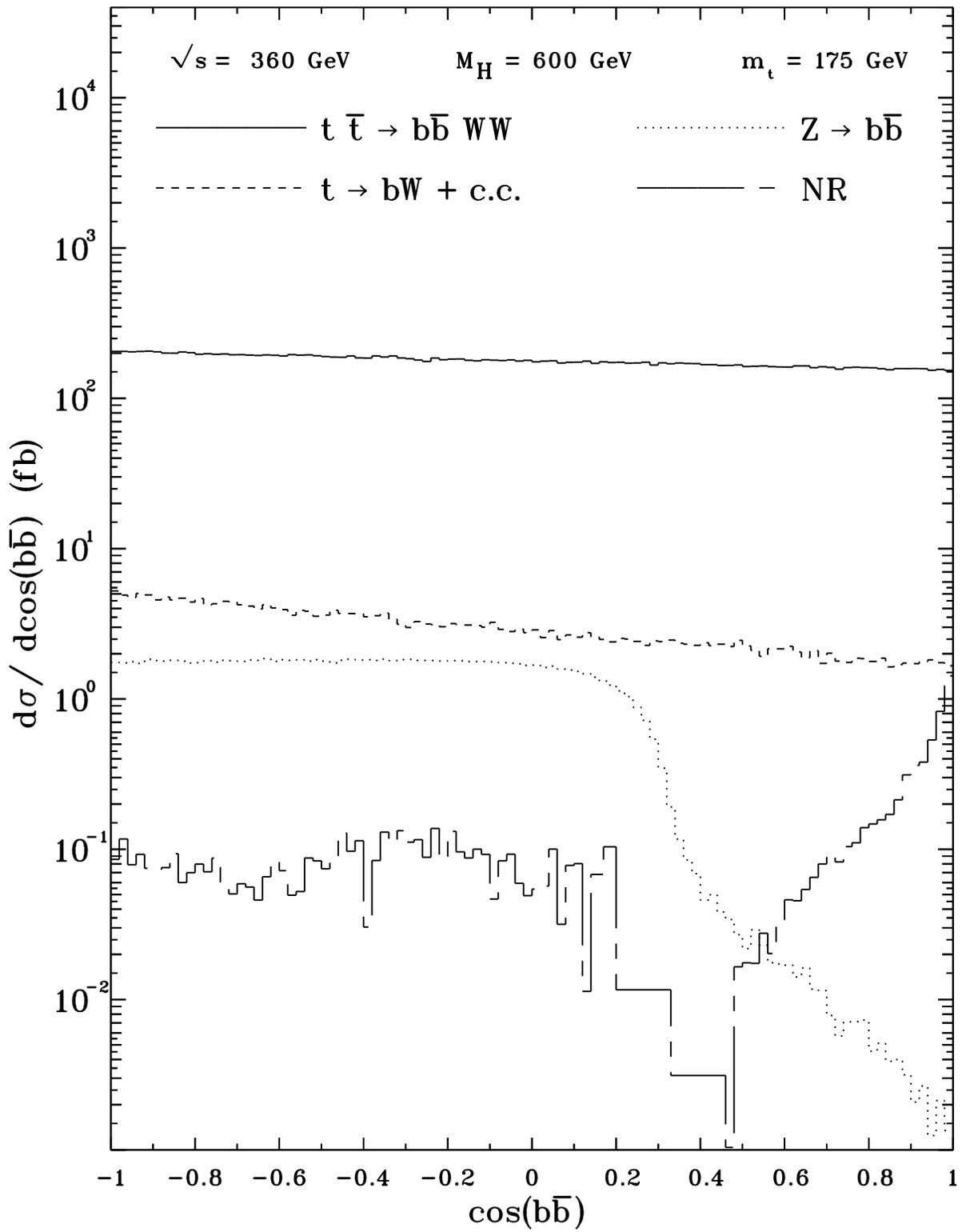
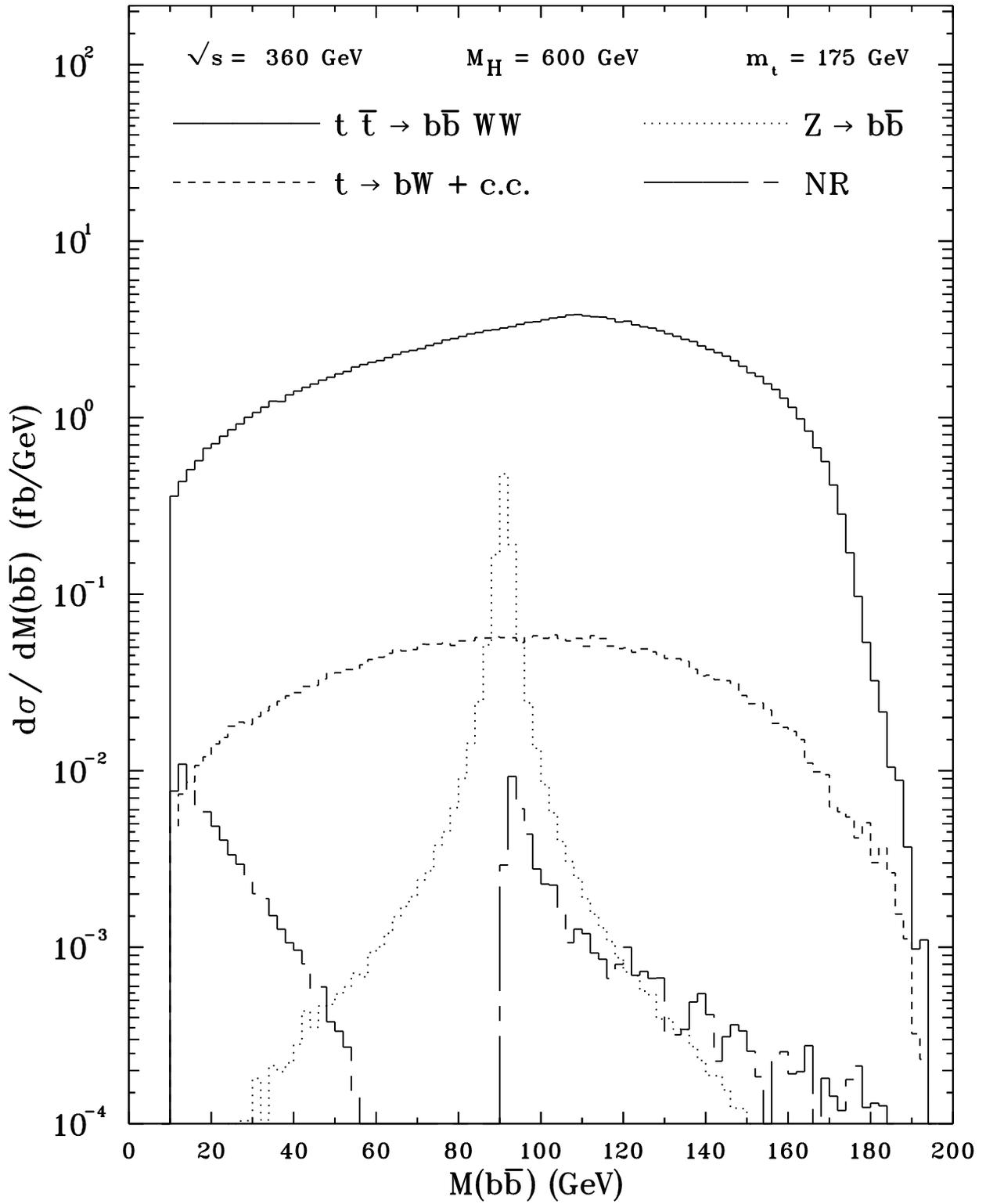
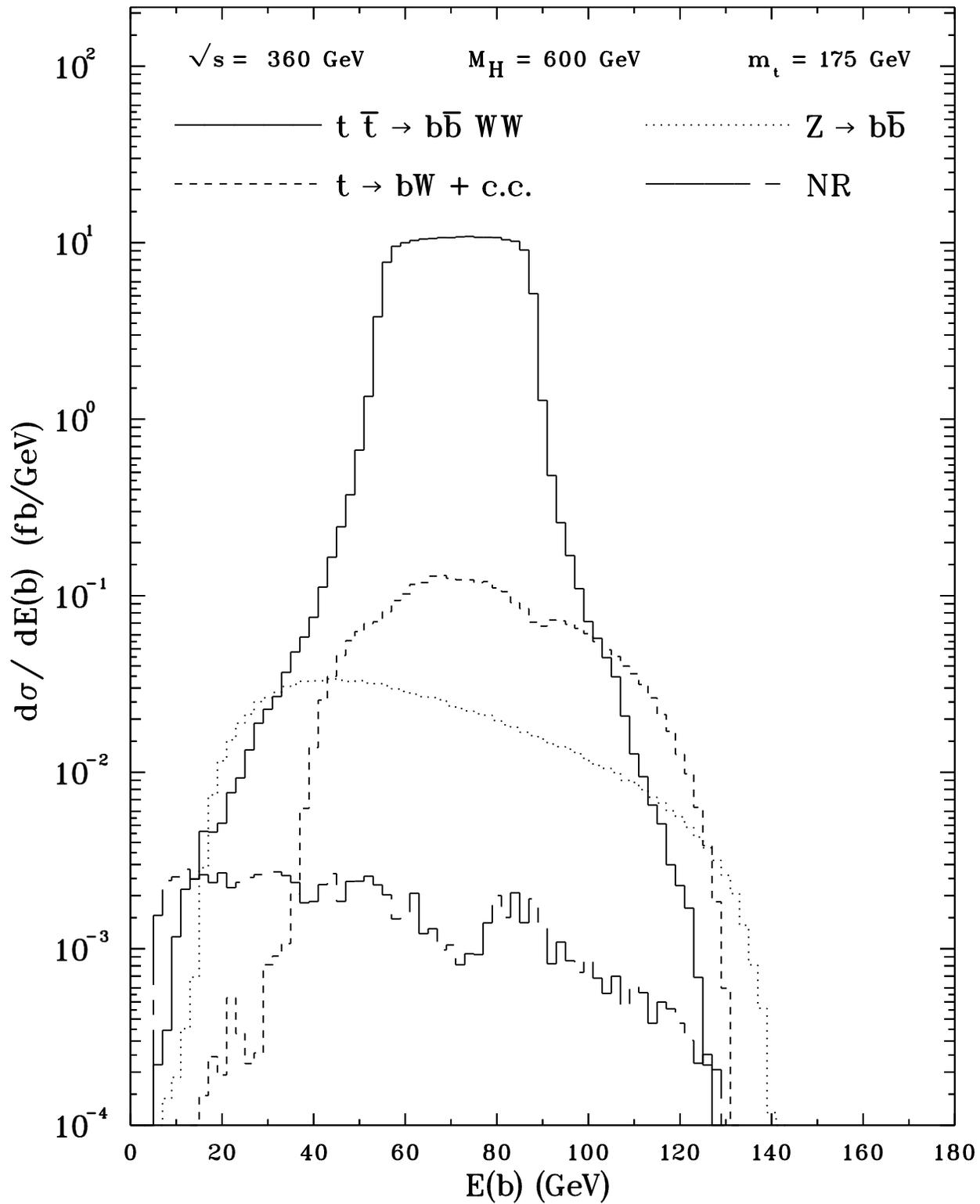
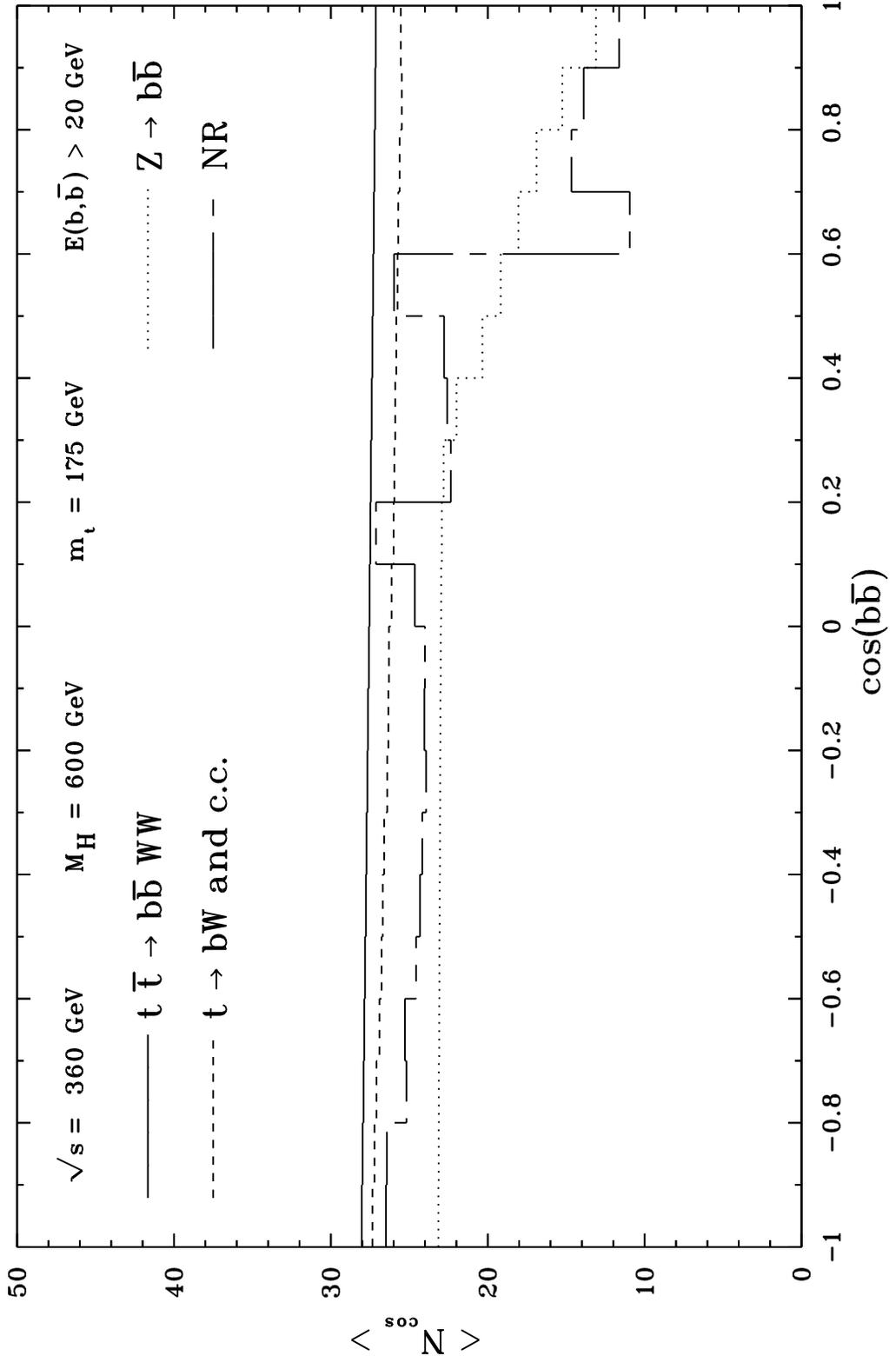
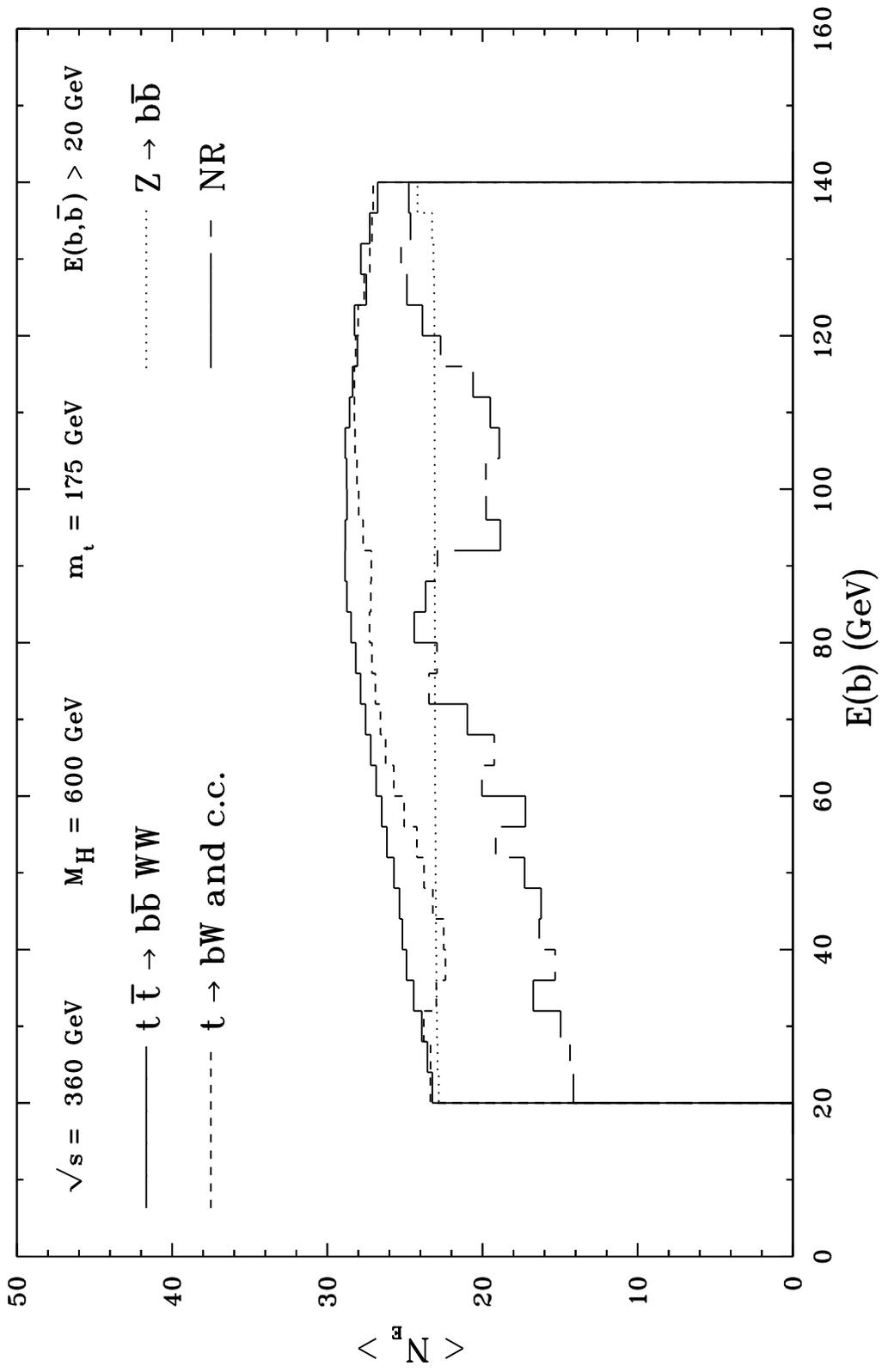
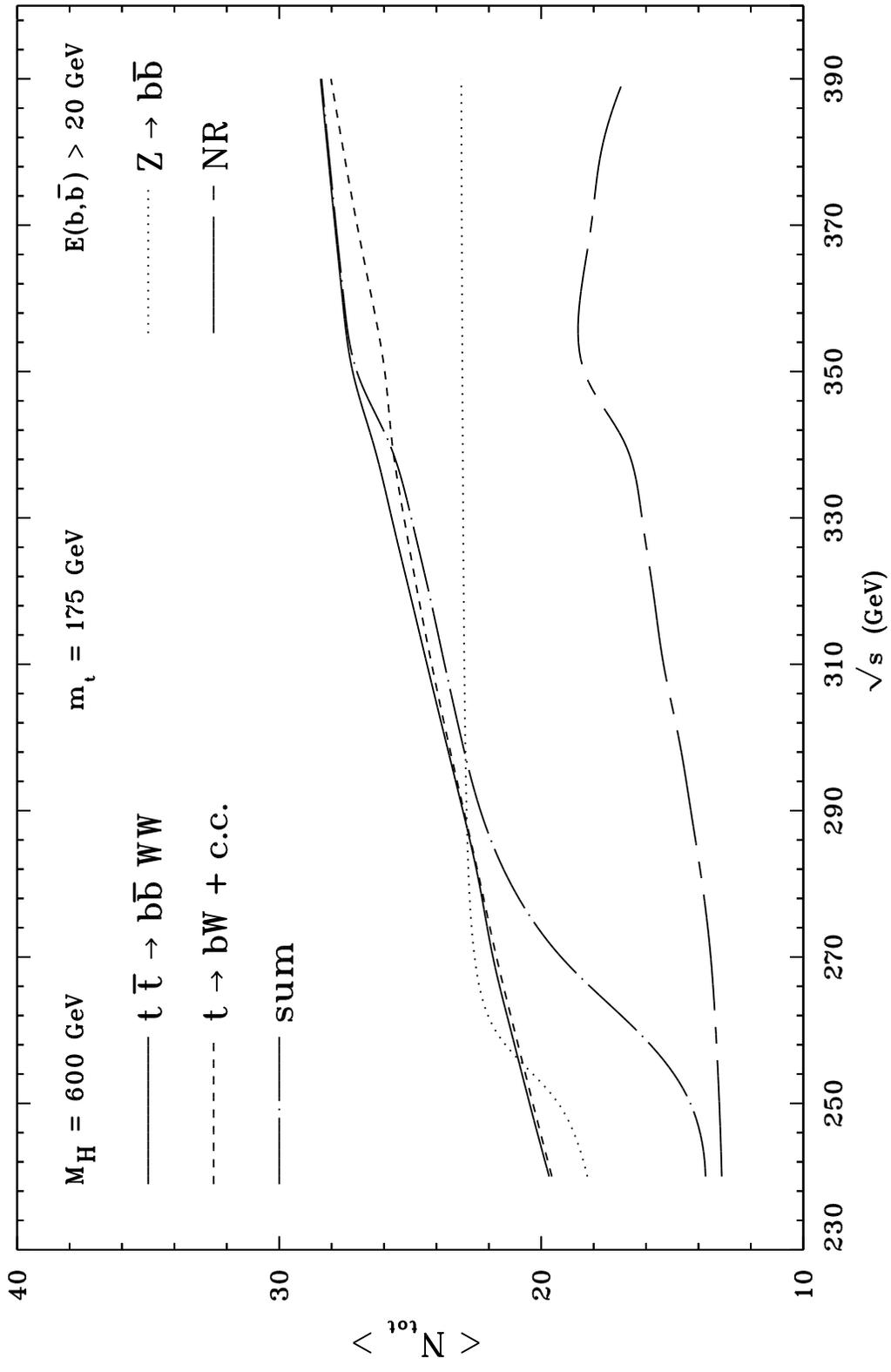
\end{document}